%% file: main.tex
\documentclass[%
 reprint,
superscriptaddress,
nofootinbib,
 amsmath,amssymb,
 aps,
prd,
longbibliography
]{revtex4-1}

\usepackage{graphicx}
\usepackage{dcolumn}
\usepackage{bm}
\usepackage{hyperref}
\usepackage{color}
\usepackage[mathlines]{lineno}
\usepackage{tabularx}
\usepackage{subfigure}
\usepackage[normalem]{ulem}
\usepackage[export]{adjustbox}
\usepackage{url}
\hypersetup{
    pdfnewwindow=true,    
    colorlinks=true,      
    linkcolor=blue,       
    citecolor=blue,       
    filecolor=blue,       
    urlcolor=blue         
}

\newcolumntype{Y}{>{\centering\arraybackslash}X}

\usepackage{natbib}
\begin{document}

\author{Kenny~C.~Y.~Ng}
\email{c.y.ng@uva.nl}
\thanks{\scriptsize \!\! \href{http://orcid.org/0000-0001-8016-2170}{orcid.org/0000-0001-8016-2170}}
\affiliation{GRAPPA Institute, University of Amsterdam, 1098 XH Amsterdam, The Netherlands}
\affiliation{Institute for Theoretical Physics, University of Amsterdam, 1098 XH Amsterdam, The Netherlands}

\author{Ariane~Dekker}
\email{a.h.dekker@uva.nl}
\thanks{\scriptsize \!\! \href{https://orcid.org/0000-0002-3831-9442}{orcid.org/0000-0002-3831-9442}}
\affiliation{GRAPPA Institute, University of Amsterdam, 1098 XH Amsterdam, The Netherlands}
\affiliation{Institute for Theoretical Physics, University of Amsterdam, 1098 XH Amsterdam, The Netherlands}

\author{Shin'ichiro~Ando}
\email{s.ando@uva.nl}
\thanks{\scriptsize \!\! \href{http://orcid.org/0000-0001-6231-7693}{orcid.org/0000-0001-6231-7693}}
\affiliation{GRAPPA Institute, University of Amsterdam, 1098 XH Amsterdam, The Netherlands}
\affiliation{Institute for Theoretical Physics, University of Amsterdam, 1098 XH Amsterdam, The Netherlands}
\affiliation{Kavli Institute for the Physics and Mathematics of the Universe~$\left( Kavli~IPMU,~WPI\right)$, University of Tokyo, Kashiwa, Chiba 277-8583, Japan}

\author{Bjarne~Bouwer}
\affiliation{Institute for Theoretical Physics, University of Amsterdam, 1098 XH Amsterdam, The Netherlands}

\author{Maurice~Geijsen}
\affiliation{Institute for Theoretical Physics, University of Amsterdam, 1098 XH Amsterdam, The Netherlands}

\author{Claudia~Glazener}
\affiliation{Institute for Theoretical Physics, University of Amsterdam, 1098 XH Amsterdam, The Netherlands}

\author{June~Groothuizen}
\affiliation{Institute for Theoretical Physics, University of Amsterdam, 1098 XH Amsterdam, The Netherlands}

\author{Jildou~Hollander}
\affiliation{Institute for Theoretical Physics, University of Amsterdam, 1098 XH Amsterdam, The Netherlands}

\author{M.~J.~F.~M.~Janssen}
\affiliation{Institute for Theoretical Physics, University of Amsterdam, 1098 XH Amsterdam, The Netherlands}

\author{Lukas~Kemme}
\affiliation{Institute for Theoretical Physics, University of Amsterdam, 1098 XH Amsterdam, The Netherlands}

\author{Wessel~Krah}
\affiliation{Institute for Theoretical Physics, University of Amsterdam, 1098 XH Amsterdam, The Netherlands}

\author{Sancho~Luijten~Perona}
\affiliation{Institute for Theoretical Physics, University of Amsterdam, 1098 XH Amsterdam, The Netherlands}

\author{Mn{\^ e}me~Stapel}
\affiliation{Institute for Theoretical Physics, University of Amsterdam, 1098 XH Amsterdam, The Netherlands}

\author{Martijn~van~Hamersveld}
\affiliation{Institute for Theoretical Physics, University of Amsterdam, 1098 XH Amsterdam, The Netherlands}

\date{July 7, 2020}

\title{Sensitivities of KM3NeT on decaying dark matter}

\begin{abstract}
The discovery of high-energy astrophysical neutrinos by IceCube has opened a new window to the Universe. However, the origin of these neutrinos is still a mystery, and some of them could be a result of dark matter interactions such as decay. Next generation gigaton water-Cherenkov neutrino telescope, KM3NeT, is expected to offer significantly improved energy resolution in the cascade channel, and advantageous viewing condition to the Galactic Center; both important for searches of dark matter decay signals. We study the sensitivity of KM3NeT on dark matter decays by performing a mock likelihood analysis for both cascade and track type events, taking into account both angular and energy information. We find that, combining both channels, KM3NeT is expected to produce world leading limits on dark matter decay lifetime in the PeV mass range, and could test some of the dark matter hints in the current IceCube data. 
\end{abstract}

\maketitle

\section{\label{sec:intro} Introduction}
\input{introduction.tex}

\section{\label{sec:formulation} Formulation}
\input{formulation.tex}

\section{\label{sec:neutrino} Neutrino Detection}

\input{neutrino.tex}

\section{\label{sec:analysis} Dark Matter Analysis}
\input{analysis.tex}

\section{\label{sec:conclusions} Conclusions and Outlook}
In this work, we consider the capabilities of KM3NeT, a gigaton water-Cherenkov neutrino telescope in the Mediterranean sea, for searching for dark matter decays from the Galactic center region.  We perform an idealized likelihood analysis that combines the cascade and track channels, and take into account the different energy and angular information for signals and backgrounds. This is especially important for the cascade channel, as KM3NeT significantly improves the angular resolution compared to IceCube. We find that KM3NeT offers significant improvements for dark matter decay sensitivity compared to the IceCube and could set world-leading limits on dark matter decay lifetime in the PeV energy range, which is competitive with gamma-ray experiments.  

In the future, this analysis can be further improved with better analysis procedure, better signal and background modeling, and also by taking into account the full angular information from all-sky observations, such as including down-going events or with a joint-analysis with IceCube.  Improved multi-messenger observations in the future will likely significantly reduce the uncertainties on both the Galactic and extragalactic astrophysical background components. These improvements could be important for probing, or even confirming signatures of dark matter decay, and potentially solving the dark matter puzzle.

\section*{\label{sec:acknowledgements} Acknowledgments}

\input{acknowledgements.tex}

\bibliography{bib.bib}

\end{document}

%% file: introduction.tex
The cold dark matter model with a cosmological constant ($\Lambda$CDM) is incredibly successful in describing many cosmological observations across various scales.  However, more than 85\% of the matter content in $\Lambda$CDM~\cite{Aghanim:2018eyx} can not be described by the Standard Model of particle physics, and it is dubbed dark matter~\cite{Bertone:2004pz}.  Understanding the properties of dark matter is one of the most pressing problems in contemporary physics. 

A popular method for dark matter identification is indirect searches of signatures of dark matter interactions such as annihilation or decay with astrophysical observations~\cite{Slatyer:2017sev, Gaskins:2016cha,Leane:2020liq}. Due to the vast landscape of dark matter models, it is important to search as many channels as possible across all energy range.  The recent rapid development of multi-messenger astrophysics has provided many great opportunities for conducting indirect dark matter searches.

Neutrino astronomy has become a reality after Icecube, a gigaton ice-Cherenkov detector at the South Pole, robustly detected extragalactic astrophysical neutrinos~\cite{Aartsen:2013jdh,Aartsen:2013bka, Aartsen:2014gkd}. Recently, there is evidence that Blazars are responsible for some of the high-energy neutrinos detected by the IceCube~\cite{IceCube:2018dnn, IceCube:2018cha}. The origin of much of the diffuse neutrino emission is still a mystery and dark matter could be responsible.

Interestingly, recent analyses of IceCube data, although not yet significant, suggest that the diffuse astrophysical neutrino spectrum may exhibit a change of spectral slope between low- and high-energy neutrino samples~\cite{Schneider:2019ayi, Stettner:2019tok}.  This could imply that there are multiple underlying neutrino sources. One possibility is neutrinos from dark matter decay~(e.g., see Refs~\cite{Esmaili:2013gha, Feldstein:2013kka, Murase:2015gea, Chianese:2017nwe, Chianese:2019kyl}), with a lifetime around $10^{28}$\,s.  More and better neutrino data as well as multi-messenger observations will be crucial to scrutinize this scenario. 

An upcoming big advancement in neutrino astronomy is the next generation gigaton water-Cherenkov neutrinos telescopes, such as KM3NeT~\cite{Adrian-Martinez:2016fdl}~(successor of Antares~\cite{ANTARES:2019svn}) and GVD~\cite{Avrorin:2019swa}. Using water as the medium brings the advantage of potentially better angular resolution for cascade events compared to ice medium due to better optical properties of the medium~\cite{Adrian-Martinez:2016fdl}. A neutrino detector in the northern hemisphere also provides a better viewing coverage of the Galactic Center, which is important for Galactic science and dark matter searches.  

Anticipating these advancements, we focus on KM3NeT and study its sensitivity to dark matter decays by considering observations towards the inner galactic region, combining both cascade and track event topologies.

%% file: formulation.tex
\subsection{Dark matter decay}
For dark matter decay, the expected signal flux is proportional to the amount of dark matter in line of sight of the detector. The dark matter mass density distribution in the Milky Way is commonly approximated by the Navarro-Frenk-White~(NFW) profile~\cite{Navarro:1996gj}, 
\begin{equation}
    \rho_{\chi}(r) = \frac{\rho_{s}}{(r/r_{s})(1+r/r_{s})^{2}} \, ,
\end{equation}
where $r$ is the Galactocentric radial coordinate, and $\rho_{s}$ and $r_{s}$ are the characteristic density and scale radius, respectively, which can be obtained by fitting the Milky-Way kinematics data.  In this work, we consider $r_s = 20$\,kpc, and normalize the profile with the local dark matter density of $0.4\,{\rm GeV/cm^{3}}$~\cite{Pato:2015dua}, where we adopt the Galactocentric radius of the solar system $R_{\odot} = 8.5$\,kpc. We choose the NFW as our canonical profile. 

To estimate the systematic uncertainty associated with the choice of density profile, we also consider the Burkert profile: 
\begin{equation}
    \rho_{\chi }(r) = \frac{\rho_0}{\left(1+{r}/{r_h}\right)\left[1+\left({r}/{r_h}\right)^2\right]}\, ,
\end{equation}
where we adopt $\rho_0$ = $1.72$\,GeV\,cm$^{-3}$ and $r_h$ = $9.26$\,kpc~\cite{Nesti:2013uwa}. The Burkert profile is a conservative choice, as the density flattens within $r_h$. We, however, note that such a large density core is inconsistent with the latest Milky-Way halo analysis~\cite{Pato:2015dua}. 
 
The differential neutrino flux from dark matter decays is computed with
\begin{equation}\label{eq:dm_flux}
    \frac{d F_{\nu} }{dE} \equiv \int d\Omega \frac{dI_{\nu}}{dE} = \frac{\Gamma}{4 \pi m_{\chi}} \frac{dN_{{\nu}}}{dE} \int_{\Delta\Omega} d\Omega \int_{0}^{\ell_{\rm max}} d\ell \rho_{\chi}\, , 
\end{equation}
where $\Gamma = 1/\tau_{\chi}$ is the dark matter decay rate, $\tau_{\chi}$ is the dark matter lifetime, $m_{\chi}$ is the dark matter mass, $dN_{{\nu}}/dE$ is the channel-dependent neutrino spectrum per decay,  $\Delta\Omega$ is angular region of interest~(ROI) considered, and $\ell$ is the line-of-sight distance, which is related to $r$ and angle between the Galactic Center and the light of sight, $\psi$, via 
\begin{equation}
    r = \sqrt{ R_{\odot}^{2} + \ell^{2} - 2 R_{\odot}\ell \cos{\psi} } \,. 
\end{equation}
The line-of-sight integral is cut off at the virial radius, $R_{\rm vir}\simeq 200$\,kpc, and thus $\ell_{\rm max} = R_{\odot}\cos{\psi} + \sqrt{ R_{\rm vir}^{2} - R_{\odot}^{2}\sin^{2}\psi}$.

\begin{figure*}
\centering
\includegraphics[width=\columnwidth]{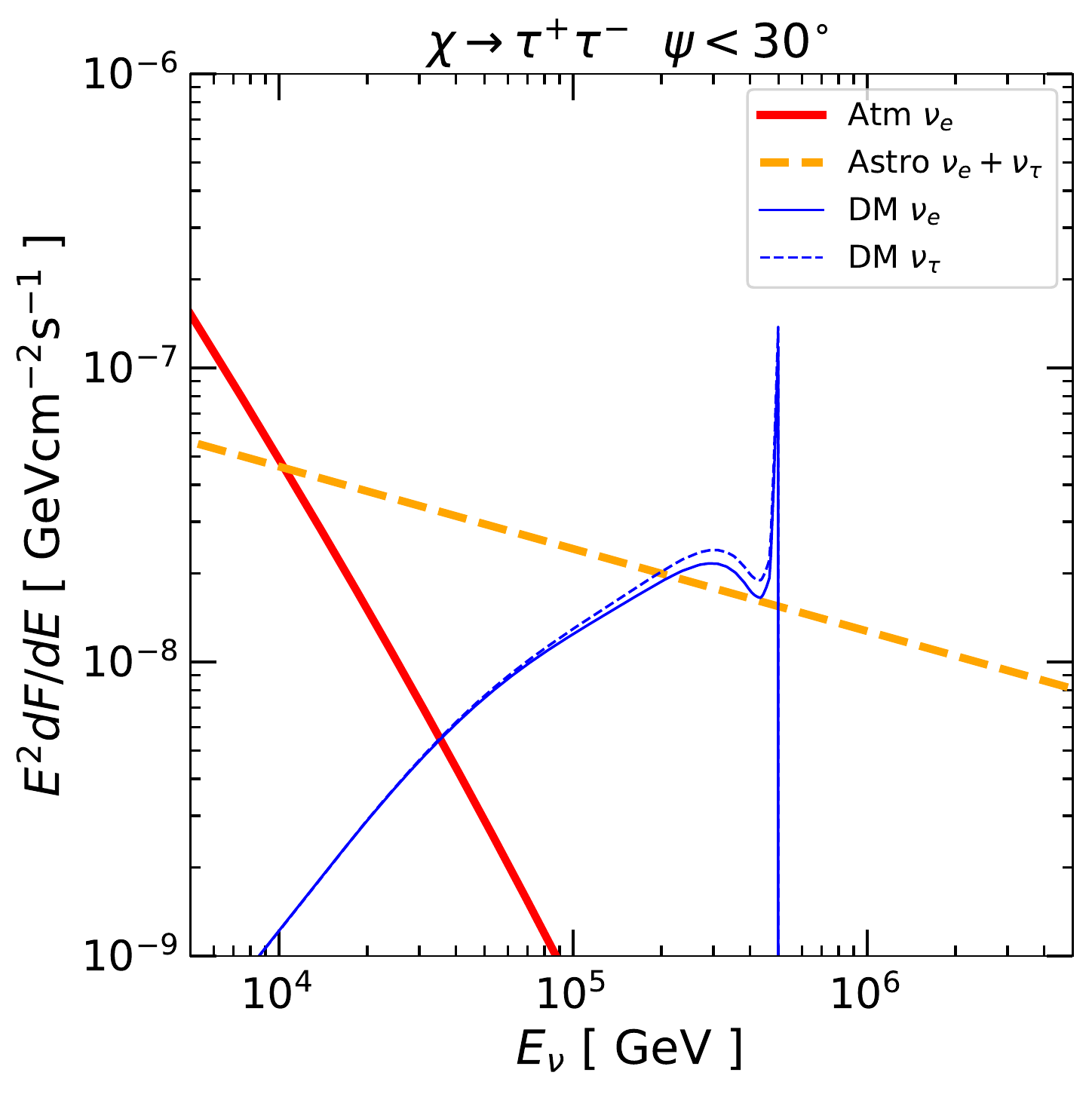}
\includegraphics[width=\columnwidth]{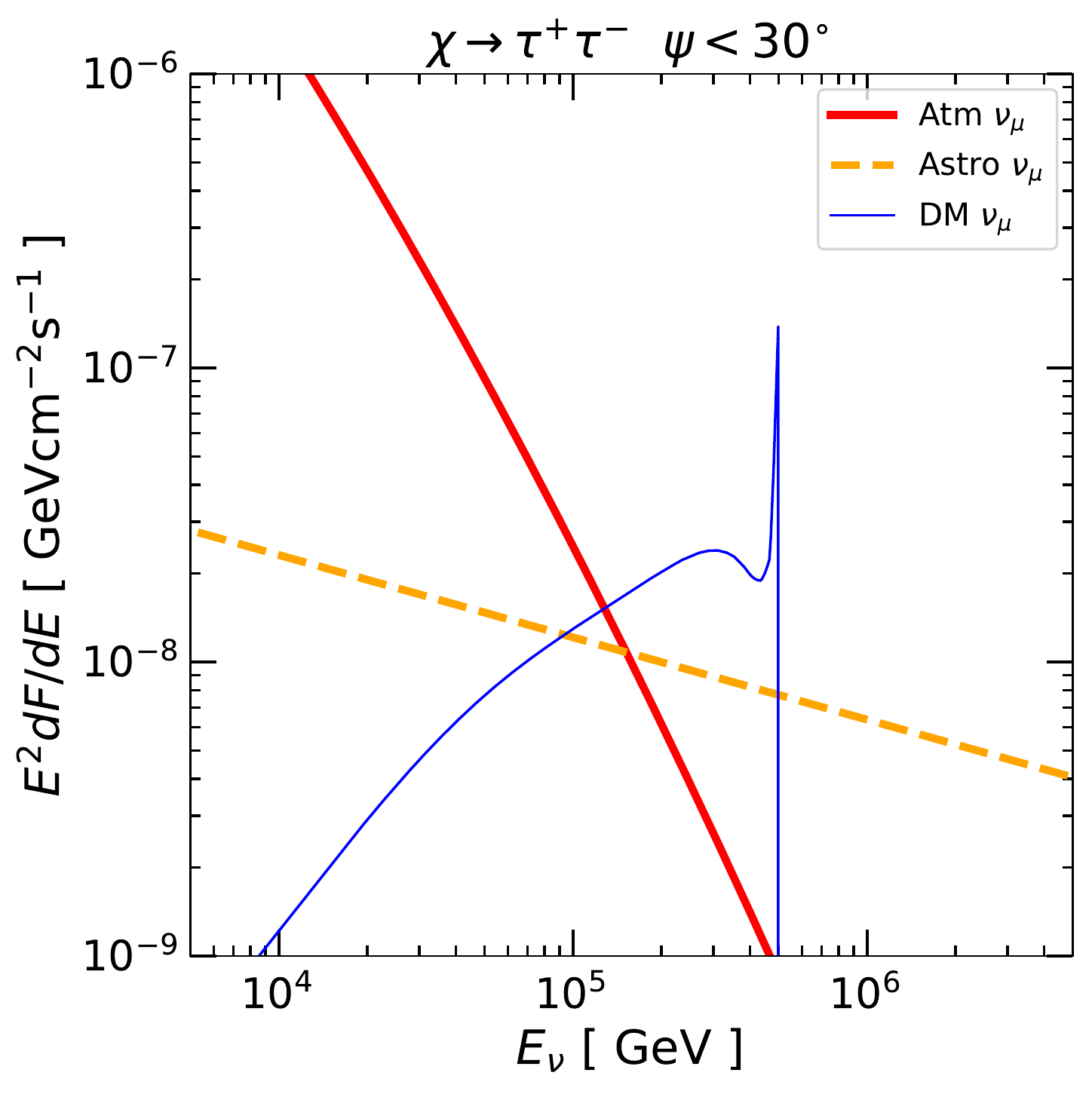}
\caption{Left: The expected neutrino flux from dark matter decays~(DM) that contributes to the cascade channel~($\nu_{e}$ and $\nu_{\tau}$). We show the case for a 1\,PeV dark matter in $\chi \rightarrow \tau^+\tau^-$ channel, with decay rate $\Gamma = 10^{-28}\,{\rm s^{-1}}$, and from the Galactic Center within an opening angle of $30^{\circ}$. Also shown are the corresponding atmospheric~($\nu_{e}$) and astrophysical~($\nu_{e} + \nu_{\tau}$) background fluxes. Right: Same as the left, but for $\nu_{\mu}$ flavor, which contributes dominantly to the tracks channel. }
\label{fig:nuflux}
\end{figure*}

\subsection{Neutrino spectra from dark matter decay}

The neutrino spectrum per dark matter decay $dN_{\nu}/dE$ depends on the dark matter mass and decay channel. 
We obtain $dN_{\nu}/dE$ using \texttt{PPPC4DMID}~\cite{Cirelli:2010xx, Ciafaloni:2010ti}, using $\chi\rightarrow b\bar{b}$ and $\chi\rightarrow \tau^+\tau^-$ to broadly represent channels with soft ($b\bar b$) and hard ($\tau^+\tau^-$) neutrino spectrum. However, we are interested in the mass range above 200\,TeV, which is beyond that computed with \texttt{PPPC4DMID}. Following Ref.~\cite{Chianese:2019kyl}, we obtain the spectra for higher dark matter masses by scaling the spectra at 200\,TeV. This approximation was found to agree well with the results from Monte Carlo generator \texttt{PYTHIA}~\cite{Chianese:2019kyl}.  

Neutrinos also change flavors during propagation~\cite{Tanabashi:2018oca}, thus the observed spectra are a mixture of the different flavor spectrum at production. The importance of this effect can be estimated by considering $E_{\delta} \equiv \delta m^{2}/L, E_{\Delta} \equiv \Delta m^{2}/L$, where $\delta m^{2} \simeq 8\times 10^{-5}\,{\rm eV^{2}}$ and $\Delta m^{2} \simeq 2\times 10^{-3}\,{\rm eV^{2}}$. We are interested in dark matter decaying in the Galactic halo, which has length scale on the order of kpc. In this case, the energy range of our interest is always much smaller than $E_{\delta}$ and $E_{\Delta}$. This means that neutrinos would quickly lose coherence and propagate as incoherent mixtures of mass eigenstates en route to the Earth. The neutrino spectrum of flavor $\alpha$ at detector is thus related to spectrum at production of flavor $\beta$ by $P_{\alpha \beta} = \sum_{i} |U_{\alpha i}|^{2}|U_{i\beta}|^{2}$, where $U_{\alpha i}$ are the elements of the PMNS mixing matrix. 

Figure~\ref{fig:nuflux} shows the three flavors of the expected neutrino flux from dark matter decay after propagation with sufficient mixing. While the neutrino spectra could be quite different between flavors at production, after propagation, the different flavors roughly average out and result in a similar spectrum. 

\subsection{Atmospheric neutrinos}

A major source of background to the dark matter signal is atmospheric neutrinos, produced by cosmic rays interacting with the Earth atmosphere. We use the all-sky averaged atmospheric neutrino flux calculated by Ref.~\cite{Honda:2015fha}, and extend it to high energies with the analytic forms provided by Ref.~\cite{Sinegovskaya:2014pia}.  Taking the typical travel distance of atmospheric neutrinos to be the radius of the Earth, $E_{\delta}$ and $E_{\Delta}$ are much smaller than the neutrino energies that we are interested in.  Therefore, we can safely ignore neutrino mixing for atmospheric neutrinos. As a result, practically there are no $\nu_{\tau}$ in the atmospheric neutrino background. 

We note that atmospheric neutrinos is more peaked at the horizon in the detector zenith angle~\cite{Lipari:2000wu, Sinegovsky:2011ab, Honda:2015fha}. But as the Galactic Center moves across different zenith angle, the sky distribution is averaged out.  For simplicity, we assumed at the end atmospheric background is isotropic in our ROI. 

Figure~\ref{fig:nuflux} shows the $\nu_{e}$ and $\nu_{\mu}$ atmospheric neutrino flux . It is worth noting that in these energy range, the atmospheric $\nu_{e}$ flux is more than one order of magnitude smaller than $\nu_{\mu}$ flux. As a result, the atmospheric neutrino background for cascades is much lower than that for tracks. 

\subsection{Astrophysical neutrinos}

For dark matter searches, the newly discovered astrophysical neutrinos is also a source of background. Although not yet significant, different analyses appear to favor different spectral indices~\cite{Stettner:2019tok, Schneider:2019ayi}. This may suggest that the underlying spectrum is not a smooth power law, and could have multiple components. 

For the purpose of estimating the sensitivity to dark matter decay, we adopt a fixed single power-law flux for the astrophysical component (per flavor)~\cite{Stettner:2019tok}, 
\begin{eqnarray}
    \frac{dI_{\nu}}{dE}&=& 1.44\times 10^{-18}\Big(\frac{E}{100\,\text{TeV}}\Big)^{-2.28}\,(\text{GeV}\,\text{cm}^{2}\,\text{s}\,\text{sr})^{-1}\, .\nonumber \\
    \label{eq:astro}
\end{eqnarray}
We assume the flavor ratio is $\nu_{e}:\nu_{\mu}:\nu_{\tau} = 1:1:1$, and it is isotropically distributed in the sky. 

Figure~\ref{fig:nuflux} shows the adopted astrophysical flux. Due to the low atmospheric $\nu_{e}$ flux, astrophysical neutrinos is a more important background component for dark matter searches in the cascade channel.

%% file: neutrino.tex
\begin{figure*}[t]
\centering
\includegraphics[width=\columnwidth]{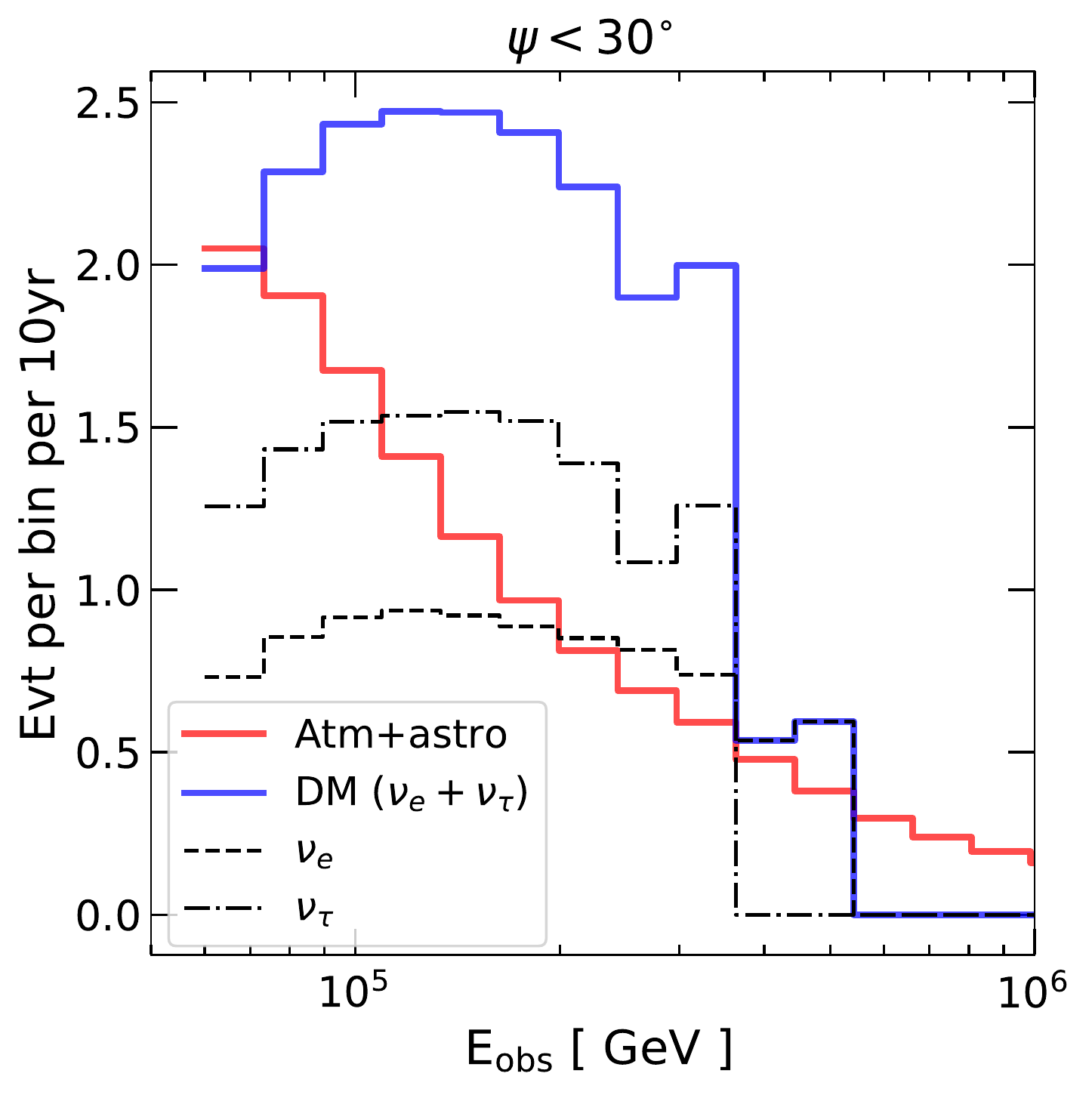}
\includegraphics[width=\columnwidth]{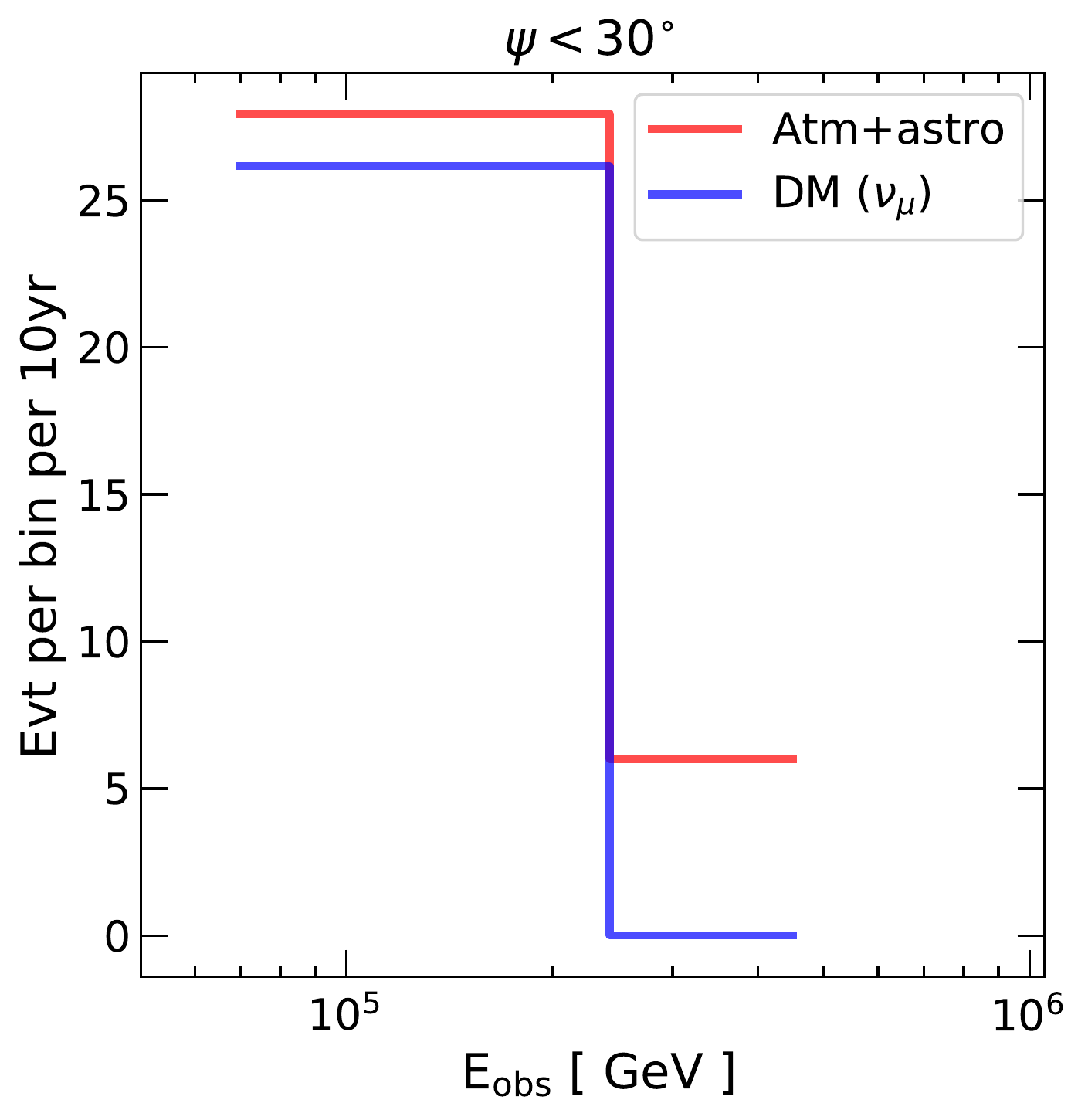}
\caption{Left: The expected number of cascade events from dark matter decay (blue) and astrophysical plus atmospheric background (red) after 10 years of KM3NeT operation from within 30$^{\circ}$ of the Galactic Center.  The parameters used are the same as Fig~\ref{fig:nuflux}. Right: Same to the left, but for tracks. Bin widths are chosen in accordance with the detector energy resolution.}
\label{fig:event_spectrum}
\end{figure*}

\subsection{KM3NeT}
We focus on the prospects of dark matter searches with KM3NeT~\cite{Adrian-Martinez:2016fdl}, which is a water-Cherenkov neutrino detector currently under construction in the Mediterranean sea.  Specifically, we consider the ARCA blocks, which consist of $\simeq 1$\,km$^3$ of instrumented detector volume after completion. 

Compared to an ice detector such as IceCube, there are two advantages for KM3NeT in the context of dark matter searches.  First, being in the northern hemisphere, the detector has a better view of the Galactic Center through the Earth, as the Earth can easily block cosmic-ray backgrounds such as energetic muons. Second, the long scattering length of Cherenkov light in water also allows improved reconstruction of cascades events, significantly improving the angular resolution of this channel. We study the DM decay sensitivity of KM3NeT taking into account both factors. Though we focus on KM3NeT, we expect the discussion also applies to GVD~\cite{Avrorin:2019swa}.

\subsection{Neutrino interactions in the detector}

We consider both cascade and track channels. Cascades are produced by electromagnetic or hadronic showers in the detector initiated by neutrino interactions including $\nu_{e}$ charged-current (CC), $\nu_{\tau}$ CC, and all-flavor neutral-current (NC) interactions. On the other hand, due to the small energy loss, the muons produced by $\nu_{\mu}$ CC interactions can travel a long distance before decaying, thus producing track-like signature in the detector. 

For NC interactions, the cross sections are smaller than that of CC, and the shower energy is a smaller fraction of the neutrino energy. We thus ignore the NC contribution in this work. We also ignore the small probability of the tau lepton decaying into a muon and producing a track event. In principle, some of the $\nu_{\tau}$-CC interactions could produce distinct event topologies~\cite{Learned:1994wg}; and hadronic and electromagnetic showers might be differentiated using ``echoes''~\cite{Li:2016kra}.  For simplicity, we neglect these potential improvements. 

For a given neutrino intensity, $dI/dE_{\nu}$, the expected number of events per bin is
\begin{equation}\label{eq:event}
N_{i,j} = \int_{i} d E_{\rm obs} \int_{j} d\Omega \frac{dI}{dE_{\nu}} \frac{dE_{\nu}}{dE_{\rm obs}}\, A_{\rm eff}\, T_{\rm eff}\, \left<e^{-\tau(E_{\nu})}\right>\, ,
\end{equation}
where $i$ and $j$ denote the energy and angular bins, respectively, $E_{\rm obs}$ is the energy deposited in the detector, $A_{\rm eff}$ is the effective area~\cite{Adrian-Martinez:2016fdl} as a function of energy, $T_{\rm eff}$ is the effective exposure time, and $\tau(E_{\nu})$ is the optical depth of the neutrinos when they travel through the Earth. 

The factor $dE_{\nu}/dE_{\rm obs}$ relates the neutrino energy to the deposited energy in the detector.  We adopt values based on analyses of IceCube events~\cite{DAmico:2017dwq}.  For $\nu_{e}$, $E_{obs}$ is the same as $E_{\nu}$. For $\nu_{\tau}$, $E_{obs}$ is slightly less than $E_{\nu}$ due to energy loss into neutrinos from $\tau$ decays.  Due to the small energy loss of muons, and their possibility of not being contained in the detector, $E_{\rm obs}$ and $E_{\nu}$ could be very different for $\nu_{\mu}$.

For neutrino detectors, atmospheric muons pose a challenging background for both cascades and tracks. Luckily, muons can be efficiently shielded by considering neutrino events from beneath the horizon. At the KM3NeT-ARCA site in the southern Italy, we track the sky position of the Galactic Center with \texttt{Astropy}~\cite{astropy:2018}, and find that it spends about 37\% of time each day below the horizon. While this factor only applies to the specific declination angle of the Galactic Center, for simplicity and due to the relatively small ROI considered, we take this factor as a constant in the calculation. Thus, we obtain the effective exposure time from the KM3NeT operation time through $T_{\rm eff} = 0.67 T$.  We consider 10 years of KM3NeT operation time for the rest of the discussion.  

For these exposures, we also include a time-averaged correction factor that takes into account neutrino absorption. The optical depth is $\tau(E_{\nu}) = n_{\oplus} \sigma(E_{\nu}) L$, where $n_{\oplus} = 3\,{\rm g\,cm^{-3}} N_{A}$ is the average density of the Earth times the Avogadro number $N_{A} = 6.02\times 10^{23}/{\rm g}$, $\sigma$ is the total neutrino-nucleon cross section~\cite{Gandhi:1998ri}, and $L$ is the time dependent cord length through the Earth towards the Galactic Center. After time averaging, the absorption factor only becomes important at high energies, for about a factor of 2 suppression at 1\,PeV. For simplicity, again, we use the same factor for the whole ROI. We also ignore the small contribution at low energies due to $\nu_{\tau}$ regeneration~\cite{Dutta:2000jv, Bugaev:2003sw}.

\begin{figure*}[t]
\centering
\includegraphics[width=\columnwidth]{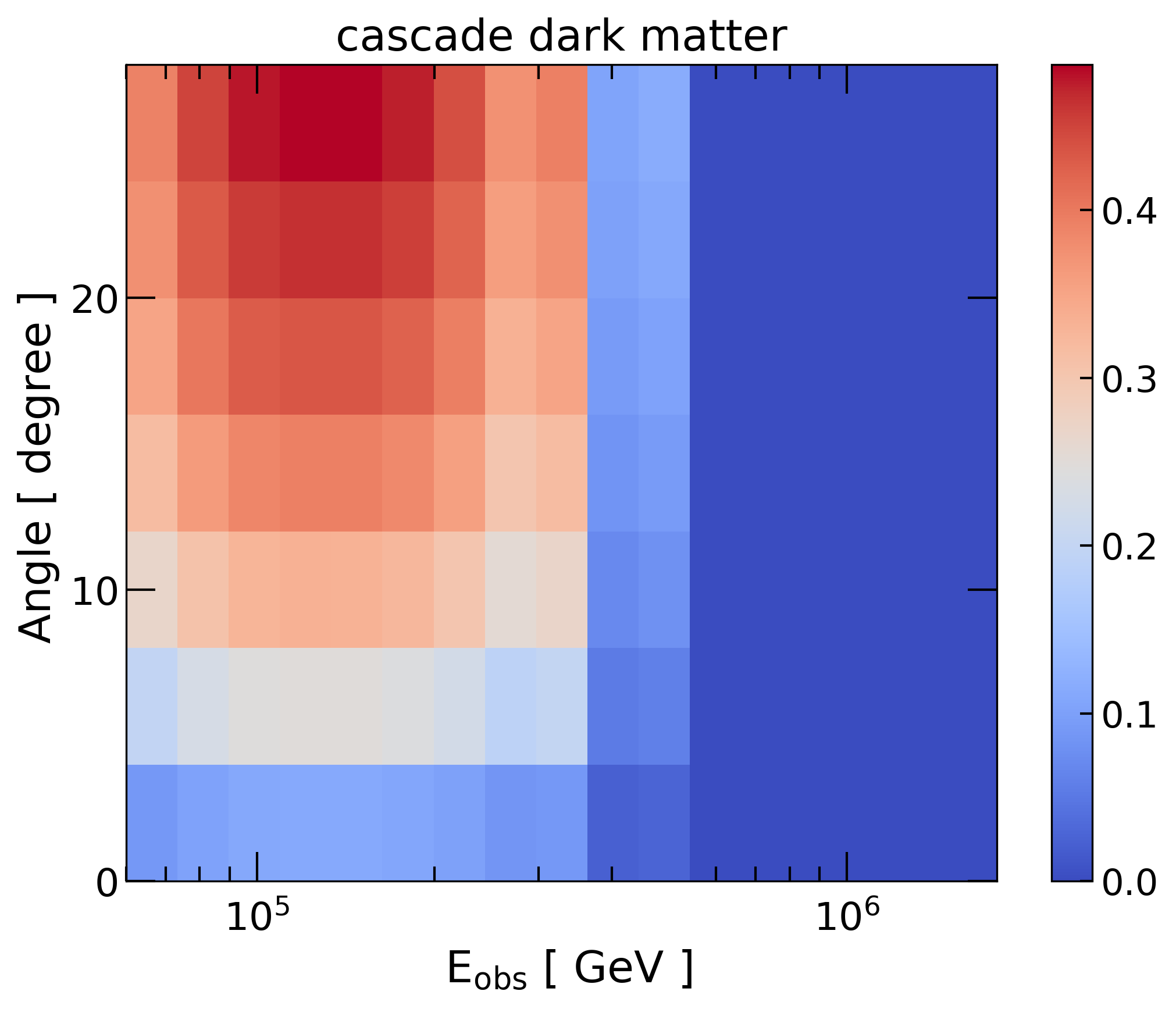}
\includegraphics[width=\columnwidth]{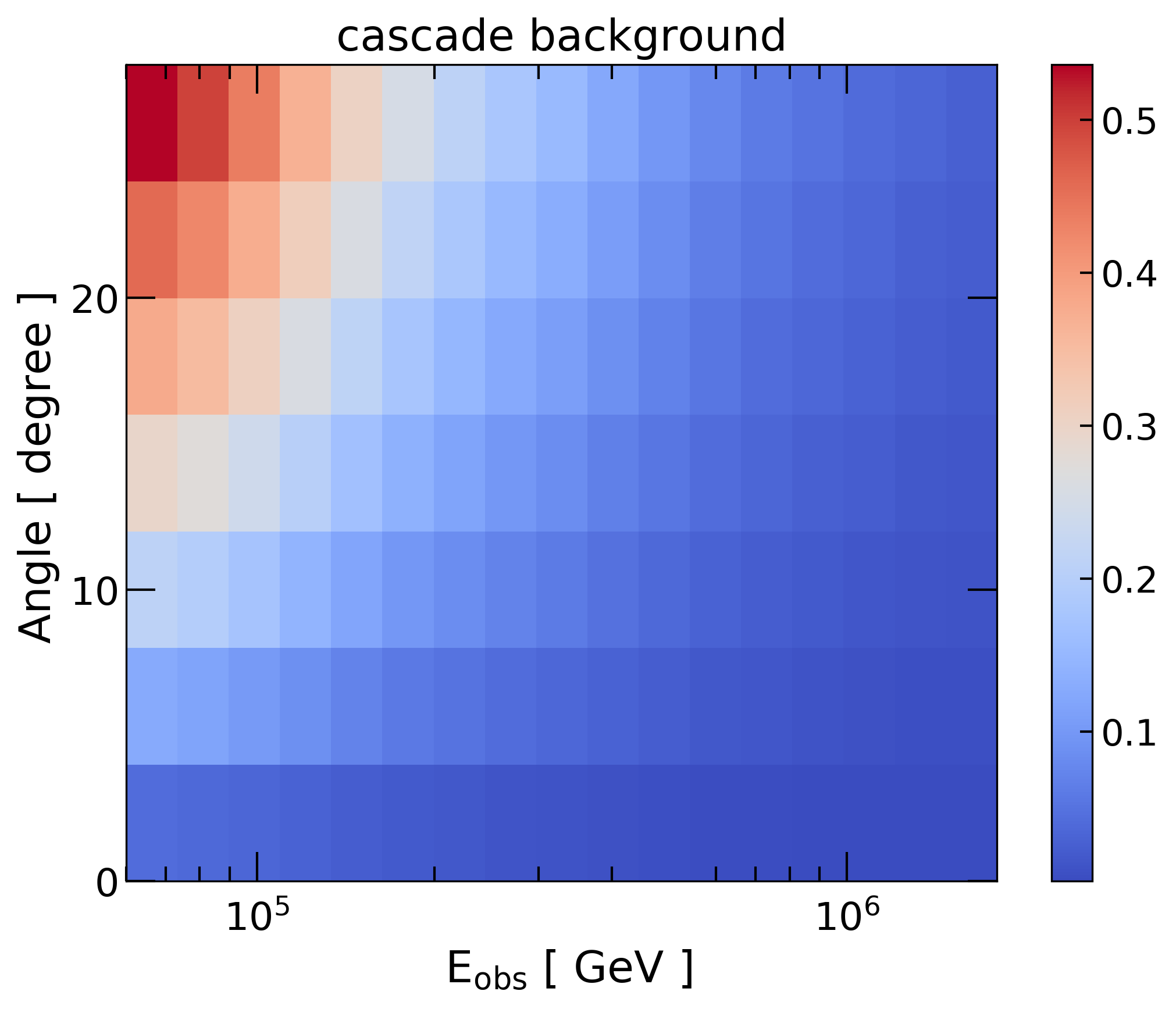}
\includegraphics[width=\columnwidth]{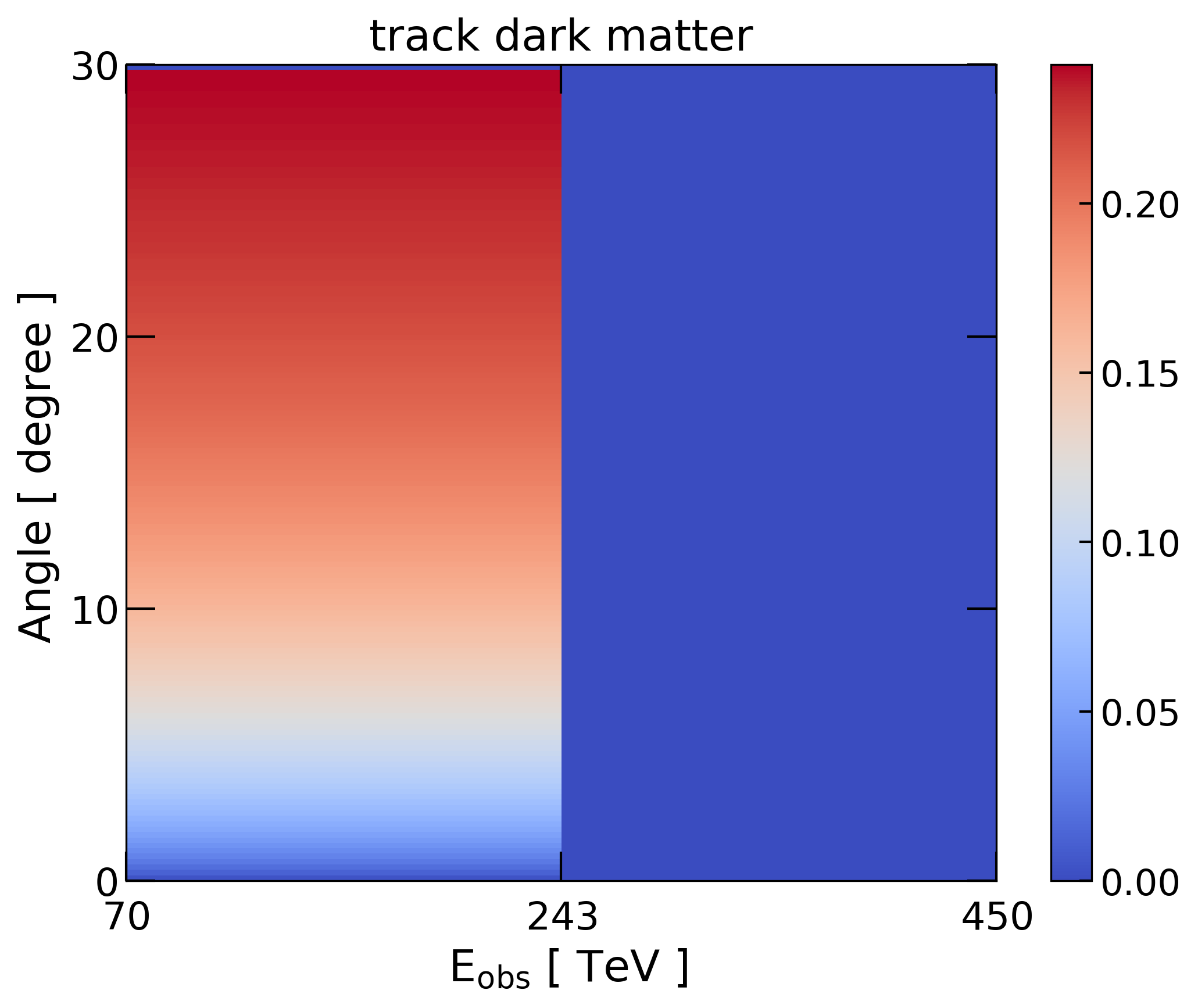}
\includegraphics[width=\columnwidth]{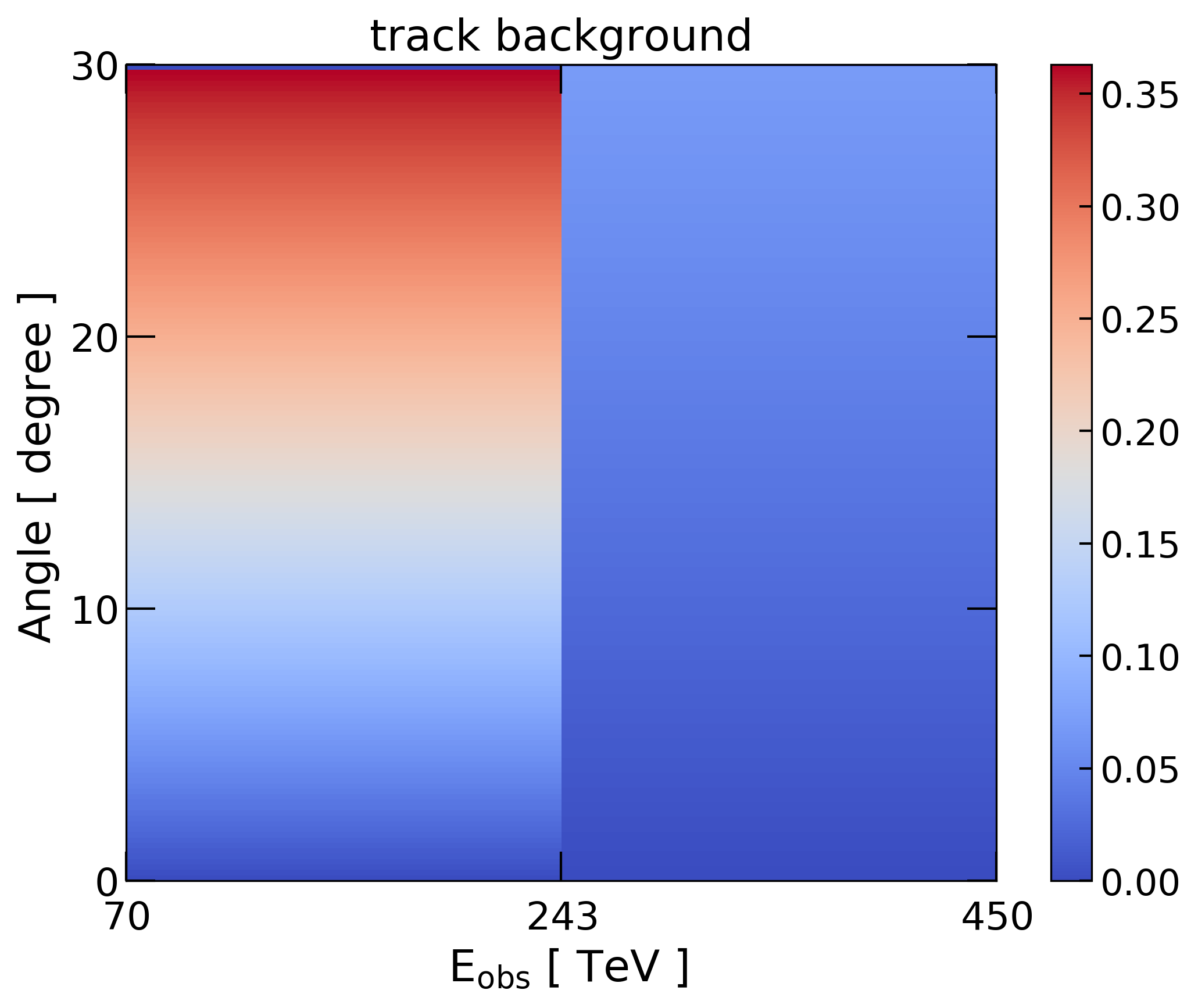}
\caption{The relative angular and energy distribution~(binned) for both dark matter decay and the background components. The parameters used are the same as Fig~\ref{fig:nuflux}. This highlights the different energy and angular distribution of dark matter versus background, both contribute to the dark matter search sensitivity.  }
\label{fig:event_pdf}
\end{figure*}

\subsection{Event distribution}

\subsubsection{Cascades}
Using Eq.~(\ref{eq:event}), we obtain the binned energy and angular distribution of the cascade events based on the energy and angular resolution of KM3NeT~\cite{Adrian-Martinez:2016fdl}.  We consider deposited energy in the range of 60--2000\,TeV, which corresponds to the same energy for $\nu_{e}$, and slightly higher for $\nu_{\tau}$. In this energy range, we bin the events according to $\Delta E/E = 0.2$ energy width.  We center the ROI at $30^{\circ}$ around the Galactic Center and divide the region into constant $4^{\circ}$-wide annuli. Both the energy and angular bins are conservatively chosen to be about twice of the energy and angular resolutions. We note that the expected angular resolution of KM3NeT cascade events is much better than that of the IceCube, for which the angular error is about $10^{\circ}$ or worse~\cite{Aartsen:2014gkd}. 

Figure~\ref{fig:event_spectrum} shows dark matter and background event spectra, using the same parameters as in Fig.~\ref{fig:nuflux}. It is clear that the final dark matter $\nu_{\tau}$ event distribution (dot-dashed; left panel) is slightly shifted down compared to $\nu_{e}$ (dashed).  With the good energy resolution, the different spectral shape between the signal and the background can been seen. 

Figure~\ref{fig:event_pdf} shows the full event distribution in both the prescribed energy and angular bins. As the background distribution is isotropic, the expected number of event simply increases with the solid angle of the annuli. The dark matter signal, however, traces the underlying dark matter column density distribution in the sky.  The plot highlights the difference of dark matter and background in both angular and energy distribution, both contributes to the dark matter sensitivity of this channel.

\subsubsection{Tracks} \label{sec:track_distribution}

For the track events, the energy resolution is much worse than the cascades due to the difficulty in reconstructing the muon energy, especially when the muon is not fully contained in the detector. For KM3NeT, the muon energy reconstruction is expected to be $\Delta \log_{10}E_{\mu} = 0.27$. Following Ref.~\cite{DAmico:2017dwq}, we consider a range of the deposited energy of 70--450\,TeV, which maps to neutrino energy between about 200 and 1000\,TeV. The poor energy resolution limits the number of energy bins that we can adopt.  Conservatively taking each bin width two times of $\Delta E_{\mu}$, we consider two independent neutrino energy bins 200--700\,TeV and 700--2000\,TeV. 

Due to long muon tracks inside the detector, the angular resolution is much better, achieving better than $0.1^{\circ}$ above 200\,TeV~\cite{Adrian-Martinez:2016fdl}. We divide the ROI into annuli with width twice of the angular resolution, which is evaluated at the central of the energy bin. This results in a bit more than 150 angular bins for the two energy bins. 

Figure~\ref{fig:event_spectrum} shows the dark matter and background event energy distribution from within $30^{\circ}$ of the Galactic center, using the same parameters as in Fig.~\ref{fig:nuflux}. With only two energy bins, the energy information provides limited discrimination power between dark matter signal and background. 

Figure~\ref{fig:event_pdf} shows the full event distribution in both the prescribed energy and angular bins from the $30^{\circ}$ ROI. While there are only two energy bins, significantly more angular bins are available.

%% file: analysis.tex
\subsection{Likelihood analysis}

\begin{figure*}
\centering
\includegraphics[width=\columnwidth]{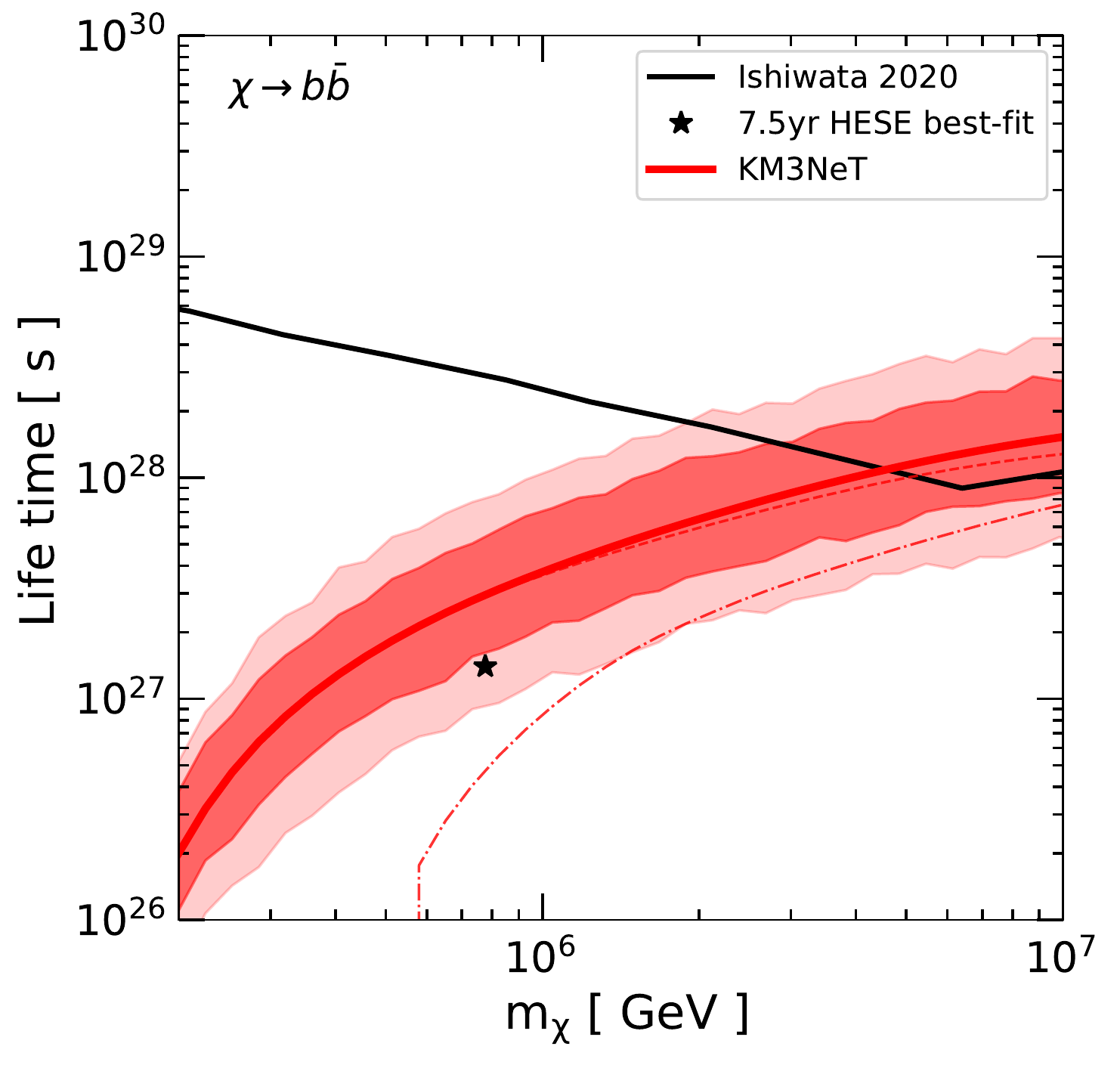}
\includegraphics[width=\columnwidth]{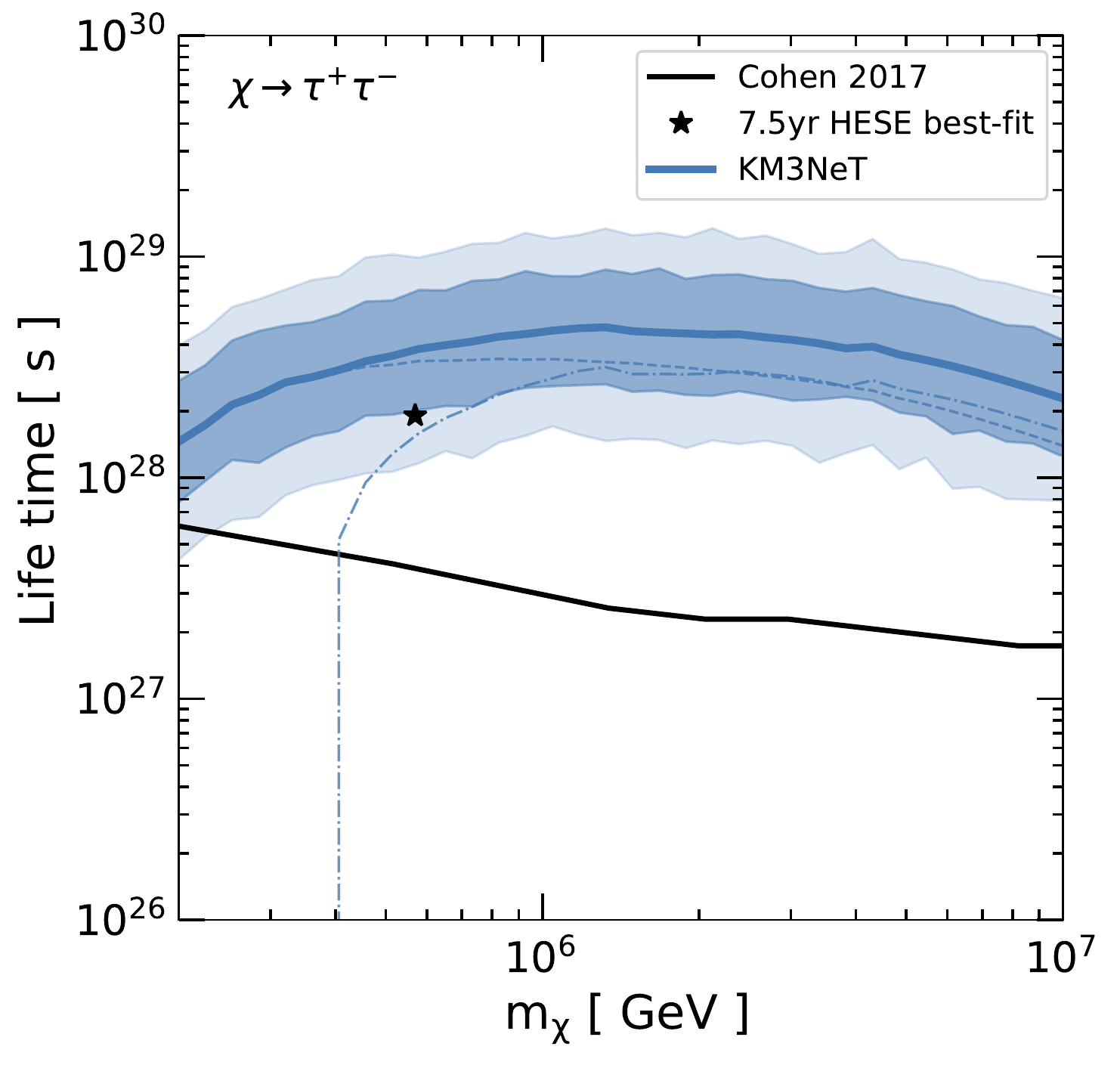}
\caption{The expected constraints to be obtained by KM3NeT after 10 years of operation for $b\bar{b}$~(left) and $\tau^{+}\tau^{-}$~(right) channels. The shaded regions correspond to the 68\% and 95\% range for the expected sensitivity obtained with Monte Carlo simulations. 
The thin dashed and dot-dashed lines corresponds to cascade-only and track-only analyses, respectively. Also shown are the best-fit parameter points for the dark matter interpretation of the 7.5~yr IceCube HESE analysis~\cite{Chianese:2019kyl}, and the strongest gamma-ray constraints from Refs.~\cite{Ishiwata:2019aet, Cohen:2016uyg}.}
\label{fig:constraint}    
\end{figure*}

We estimate the sensitivity of KM3NeT to dark matter decays by performing a simple likelihood analysis on mock data sets. For each dark matter mass and decay channel, we consider the following likelihood function:
\begin{equation}
    \mathcal L(\Gamma) = \prod_k {\cal{L}}_{k}\left[n_{k}|\mu_{k}\left(\Gamma\right) \right], 
\end{equation}
where $\mu_{k}(\Gamma)$ is the expected number of events per bin given by the sum of the signal and background contributions, $n_{k}$ is the number of events in the mock data set. ${\cal{L}}_{k}$ is the Poisson likelihood of observing $n_{k}$ events given the expectation $\mu_{k}$, and the index $k$ runs over all the angular and energy bins for both cascade and track channels.

For each dark matter mass and decay channel, we generate 500 mock data sets of event bins drawn from the Poisson distribution of the expected background-only event prediction~($\mu_{k}(0)$). We then compute the test statistic~(TS) as
\begin{eqnarray}
    {\rm TS}(\Gamma) &=& -2\ln\left[\frac{\mathcal L(\Gamma)}{\mathcal L(0)}\right]\, .
    \label{eq:TS}
\end{eqnarray}
The 95\% one-sided confidence level upper limit on $\Gamma$ is then obtained with ${\rm TS}(\Gamma_{95}) = 2.71$. Running over all the mock data sets, we then obtain the mean expected dark matter limit and their variance. 

\subsection{Results}
Figure~\ref{fig:constraint} shows the mean expected dark matter constraints and its range for $\chi \rightarrow b\bar{b}$ and $\chi \rightarrow \tau^{+}{\tau^{-}}$, respectively. We show both the results obtained from cascades and tracks, as well as the combined sensitivity. We can see that in general, the cascade channel is expected to be more sensitive than the tracks channel, despite having a smaller effective area. At lower dark matter masses, the track sensitivity drops sharply. This is because the higher neutrino energy threshold compared to cascade events~(see Sec.~\ref{sec:track_distribution}). 

We note that our results are relatively insensitive to the density profile choice. If we instead use the flat Burkert profile, the sensitivity only decreases by about 20\%. This is because of having only one power of density in the line-of-sight integral in Eq.~\ref{eq:dm_flux} (as opposed to two powers for dark matter annihilation).

For comparison, we also show the best-fit parameter regions from Ref.~\cite{Chianese:2019kyl}, which considered the the 7.5-yr IceCube HESE events and interpreted some of the events as potential dark matter signals. We can see that KM3NeT is expected to have the sensitivity to test these parameter space. 

Compared to existing limits obtained with IceCube data~\cite{Aartsen:2018mxl, Chianese:2019kyl}, our sensitivity is generally a factor of a few to about one order of magnitude better.  This is a result of the larger column density towards the Galactic Center, the larger effective area in KM3NeT (according to its letter of intent~\cite{Adrian-Martinez:2016fdl}) compared to IceCube HESE analysis, and the improved angular resolution of the cascade channels. We note that in Ref.~\cite{Aartsen:2018mxl}, the $b\bar{b}$ limit improves rapidly when dark matter mass becomes larger than 1\,PeV, which is attributed to enhanced low-energy neutrino tail due to QCD fragmentation. Taking such effect into account is beyond the scope of this work, but we expect that KM3NeT would out-perform IceCube for Galactic dark matter decay searches in general, once such effect is taken into account.  

Most of the dark matter decay channel produces neutrinos as well as gamma rays, which also offers a powerful avenue for complementary dark matter searches~\cite{Murase:2012xs, Murase:2015gea}. We thus also compare with the strongest constraints obtained with various gamma-ray data sets~\cite{Cohen:2016uyg, Abeysekara:2017jxs, Ishiwata:2019aet}. For $b\bar{b}$, we find the strongest constraint from Ref.~\cite{Ishiwata:2019aet}. In this case, the gamma-ray constraint is stronger at lower energies, but KM3NeT is expected to overtake above a few PeV~(before considering the boost from \cite{Aartsen:2018mxl}).  For $\tau^{+}\tau^{-}$, KM3NeT is expected to out-perform gamma-ray probes and set world-leading constraints. 

Another powerful way for dark matter searches is the angular power spectrum analysis that utilizes full-sky angular information. Compared with the $\tau^{+}\tau^{-}$ sensitivity obtained from Ref.~\cite{Dekker:2019gpe} (tracks only), our combined sensitivity is slightly better. 

\subsection{Discussions}
We have performed a simplified binned-likelihood analysis to estimate the dark matter decay sensitivity of gigaton water-Cherenkov neutrino detectors, such as KM3NeT. There are several factors that could be considered in the future, which could offer further improvements. 

For simplicity, we choose to perform a binned analysis, consider a relatively small sky portion, and conservatively taking the bin width twice the energy and angular resolutions.  In principle, an unbinned analysis would offer the most complete information on the events. All-sky observations including above horizon events can also be combined statistically. Adding these information should further improve the sensitivity. 

We have assumed all $\nu_{\tau}$ events would produce cascade signatures in the detector and is indistinguishable with $\nu_{e}$ cascade events.  Future analyses with improved flavor selection capabilities~\cite{Learned:1994wg, Li:2016kra, Wille:2019pub} could improve the signal-to-noise of dark matter searches, as the atmospheric neutrino background is practically $\nu_{\tau}$-free above $\sim$\,TeV. 

The predicted dark matter signal depends on the neutrino spectrum per dark matter decay.  For heavy dark matter, the spectrum could be significantly modified by high-order effects.  We have taken a simple scaling approach in our spectrum modeling, which were test against \texttt{Pythia}~\cite{Ciafaloni:2010ti,Chianese:2019kyl}.  However, significant corrections could still be present that were not covered by \texttt{Pythia}~\cite{Aartsen:2018mxl, Ishiwata:2019aet}, which will affect the interpretation of the dark matter parameter space. 

Our analysis have assumed a fixed single-power law astrophysical background model. In practice, uncertainties on the astrophysical component could affect the dark matter analysis, which could be incorporated into the analysis, where the background models are fit to the data continuously. If the astrophysical component is more complicated than expected, such as having more than one spectral components, the dark matter sensitivity could weaken depending on the assumed background model in the analysis due to potential degeneracies. In that case, improved neutrino source identification and multi-messenger observations will be important to minimize the background model uncertainty. 

For simplicity we ignore the extragalactic dark matter component in this work, which also makes our results conservative. Towards the Galactic Center the extragalactic contribution is subdominant, but for a full-sky analysis it is not, and should be considered in the future. 

One advantage of KM3NeT dark matter analysis is its ability to observe the Galactic Center, which has a larger dark matter column density. The Galactic Center, however, could also be a source of astrophysical neutrinos in additional to the extragalactic component. This component, if exist, could be difficult to distinguish from dark matter.  Future high-energy gamma rays instruments, such as the  Cherenkov Telescope Array\footnote{https://www.cta-observatory.org/} and the Southern Wide-Field Gamma-Ray Observatory~\cite{Abreu:2019ahw} could be used to constrain the Galactic source contribution.  

%% file: acknowledgements.tex
We thank Rasa Muller, Aart Heijboer, and especially Marco Chianese and Suzan Du Pree for useful comments and suggestions. We acknowledge the software packages used, such as the \texttt{SciPy} ecosystem~\cite{scipy} that includes \texttt{Matplotlib} and \texttt{NumPy}; and \texttt{Astropy}~\cite{astropy:2018}. This work was supported by the European Union's Horizon 2020 research and innovation programme under the Marie Sk'{\l}odowska-Curie grant agreement No 844664 (KCYN),  JSPS/MEXT KAKENHI Grant Numbers JP17H04836, JP18H04340, JP18H04578 (SA), and the University of Amsterdam (AD and SA).

This project has been carried out in the context of the ``ITFA Workshop'' course, which is part of the joint bachelor programme in Physics and Astronomy of the University of Amsterdam and the Vrije Universiteit Amsterdam, for bachelor students (BB, MG, CG, JG, JH, MJFMJ, LK, WK, SLP, MS, MvH), supervised by KCYN, AD, and SA, following the previous successes~\cite{Aalberts:2018obr, Ando:2019rvr}. The actual work was done in three independent groups A--C during a four-week period of January 2020. Group A (CG, JH, MJFMJ, and WK) worked on the neutrino-flux computation from dark matter decay, and the likelihood analysis. Group B (BB, MG, and LK) worked on the backgrounds and the detector event computation for cascade events. Group C (JG, SLP, MS, and MvH) worked on the backgrounds and the detector event computation for track events. All the group contributed writings and figures for their corresponding sections. 
The entire project was coordinated by the supervisors in group X (KCYN, AD, SA), who collected and checked all the numerical codes developed by the students, added the Monte Carlo simulations, and brought the figures and text to refinement after the course had been completed.